# Spin polarization of electrons in quantum wires

A.A. Vasilchenko

*Kuban State Technological University, Krasnodar, Russia*

The total energy of a quasi-one-dimensional electron system is calculated using density functional theory. It is shown that spontaneous ferromagnetic state in quantum wire occurs at low one-dimensional electron density. The critical electron density below which electrons are in spin-polarized state is estimated analytically.

The low-dimensional electron systems in semiconductors are one of the most actual and intensively developing areas in condensed matter physics for the last decades. The phenomena as integer and fractional quantum Hall effects, Wigner crystallization, metal-dielectric transition, conductivity quantization, persistent current oscillations, Coulomb blockade have been discovered in the low-dimension electron systems. The quantum wires (QW) properties have been studied less than properties of quantum wells and quantum dots. The phenomena like Wigner crystallization, spontaneous electron polarization in the zero magnetic field, "0.7 anomaly" of conductivity may take place in QW. These phenomena are still very far from a complete theoretical explanation.

The study of electron spin polarization in quasi-one-dimensional channels in zero magnetic field is of special interest. The results of experiments [1] point out on presence of a spin polarized state in a narrow QW. Ferromagnetic state in QW may take place due to exchange and correlation interaction of electrons. Previously effects of electron spin polarization in QW have been studied using Hartree-Fock method [2, 3], one-dimensional Hubbard model [4, 5], density functional theory [6, 7]. At present, the density functional theory is considered as one of the most powerful methods (excluding the exact many-body Hamiltonian diagonalization) take into account of electron-electron interaction.

In this work the density functional theory have been used to study transition of electrons in QW to the spin-polarized state in zero magnetic field.

Let us consider a single quasi-one-dimensional channel. In the direction of the $z$ axis an electron density is given by $\delta$−function, along the $x$ axis movement of electrons is quantized, and along the *y* axis electrons can move free. Inside a QW electrons are confined by a positively charged background with the two-dimensional density $n_p$ ($n_p$ differs from zero at $|x| \leq a/2$, where *a* is width of QW).

According to the density functional theory the total energy of the electron system is the functional of the electron density *n(x)*:

$$E[n] = T[n] + \frac{1}{2}\int V_H(x)[n(x) - n_p]dx + E_{xc}[n], \qquad (1)$$

where *T[n]* is kinetic energy of non-interacting electrons, $E_{xc}[n]$ is exchange-correlation energy. The second term in Eq. (1) is a Coulomb energy of electrons.

Below we use the atomic system of units, in which energy is represented in units of $Ry = \frac{e^2}{2ka_B}$, and a length in Bohr's radius $a_B = \frac{k\hbar^2}{m_e e^2}$, where $m_e$ – effective mass of an electron, *k* – dielectric constant.

As a rule, the correlation energy can be neglected, and only exchange energy is taken into account in calculations [8]:

$$\varepsilon_x(n) = -16n^{1/2}/3\pi^{1/2}g_s^{1/2}, \qquad (2)$$

where $g_s$ is spin factor.

From Eq. (1) we obtain a Kohn-Sham equations

$$-\frac{d^2\psi_i(x)}{dx^2} + V_{eff}(x)\psi_i(x) = E_i\psi_i(x), \qquad (3)$$

where

$$V_{eff}(x) = V_H(x) + V_x(x), \qquad (4)$$

$$V_H(x) = 4\int_{-\infty}^{\infty}(n_p - n(x_1))\ln|x - x_1|dx_1, \qquad (5)$$

$$V_x(x) = \frac{d(\varepsilon_x(n)\,n)}{dn}. \qquad (6)$$

Further we consider that only the lower energy level is populated, then the electron density is defined by

$$n(x) = N\psi_0^2(x), \qquad (7)$$

where $N = n_p\,a$ is linear density of electrons.

We have developed and realized an efficient algorithm for numerical solution of the Schrödinger equation derived from the Kohn-Sham system of equation (2)–(7). The algorithm has been tested. Test results show that the developed algorithm is more efficient compared with the standard diagonalization methods. The computer program of implementation of the developed algorithm to solve Kohn-Sham nonlinear equations allows to perform computations for a QW with low electron densities (small quantum well with one bound state), that is extremely important for studying of transition to the spin-polarized state and a system with strongly inhomogeneous electron gas.

Kohn-Sham nonlinear system of equations has been solved numerically using iterations method. Calculations results of the potentials and the wave function of electrons are shown in Fig. 1. It is seen that Coulomb potential contribution to the effective potential is much less, than the exchange potential.

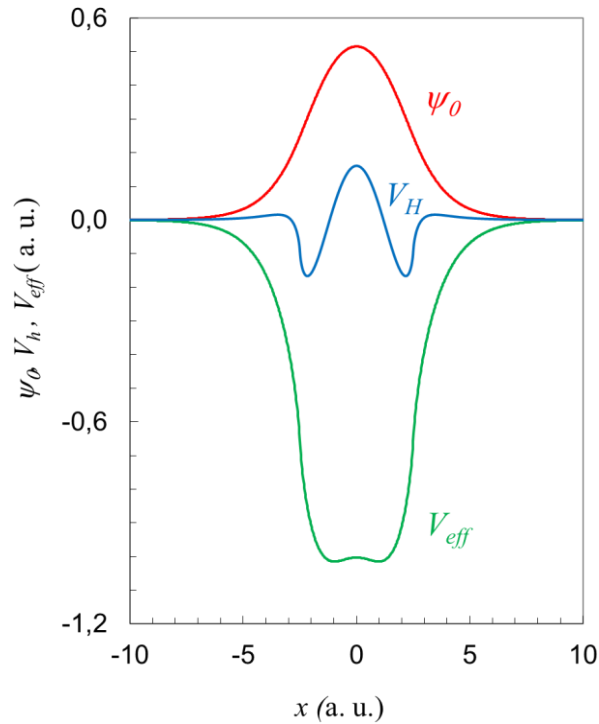

FIG. 1. Wave function, Coulomb and effective potentials along the *x* direction in a QW for *a*=5, $n_+$=0.05, $g_s$=1.

Transition to the spin-polarized state occurs at low electron densities, when only the lowest energy level $E_0$ (for not very big *a*) is populated. In this case kinetic energy has the form

$$T = \frac{\pi^2}{3g_s^2} N^3 + N(E_0 - \int V_{eff}(x)\psi_0^2(x)dx). \qquad (8)$$

From Eqs. (2) and (8) it is seen that at high densities $N$ a state with unpolarized electrons ($g_s = 2$) is always energetically favorable.

The total energy of electrons for $g_s = 1$ and $g_s = 2$ have been calculated. The results of calculations show that the state with completely polarized electrons is energetically favorable at densities $n_p < n_c$ or $N < N_c$ (Fig. 2). Note, that for narrow QW the critical density $n_c$ can be high, but the value of $N_c$ is changed weakly. From the results presented in Fig. 2 we can also estimate a magnitude of a critical density for two-dimensional case. At $d = 7$ we have the value of $n_c \approx 0{,}04$ (approximately $4 10^{10}$ cm$^{-2}$ for GaAs), which corresponds to the upper boundary of transition into the spin-polarized state of the two-dimensional electron gas.

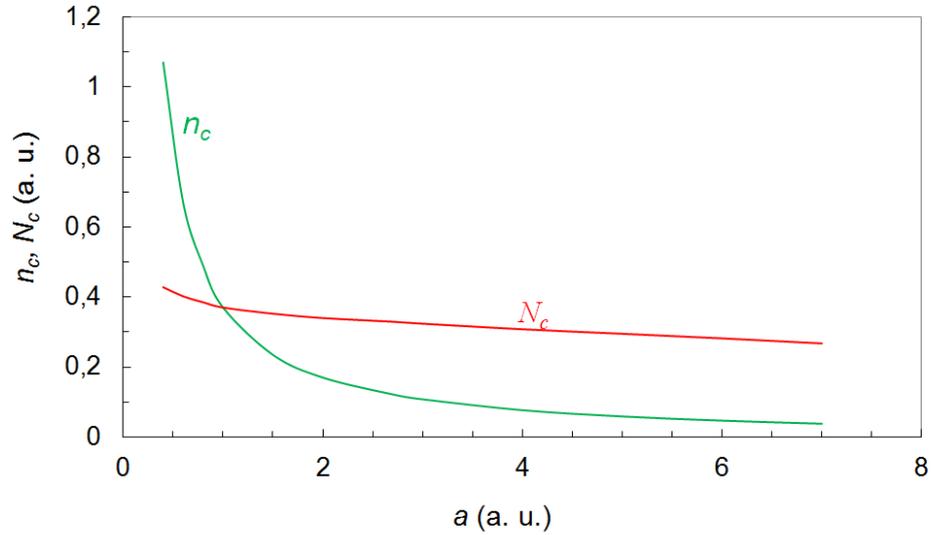

FIG. 2. The phase diagram of transition to the spin-polarized state in a QW.

Results of the calculations showed that a contribution of Coulomb interaction into a total energy is much less than a contribution of kinetic and exchange energies. Neglecting the Coulomb contribution, we estimate the total energy of a system. We take the trial wave function of electrons as

$$\psi_0(x) = \frac{\exp(-x^2/2b^2)}{\pi^{1/4}(b)^{1/2}}.$$

Then the total energy has the form

$$E = \frac{\pi^2 N^3}{3g_s^2} + \frac{N}{2b^2} - \left(\frac{2}{3}\right)^{3/2} \frac{8N^{3/2}}{\sqrt{b g_s} \pi^{3/4}}. \qquad (9)$$

The first and the second terms correspond to kinetic energy, the third term corresponds to exchange energy. From Eq. (9) we find the value of $b$, at which a minimum of the total energy is achieved:

$$b = \frac{3\pi^{1/2}}{4} \left(\frac{g_s}{2N}\right)^{1/3}. \qquad (10)$$

Using Eqs. (9) and (10) we obtain for the value of the critical density $N_c = 0.3$. This value correlates well with exact values presented in Fig. 2.

We can also estimate the value of the critical density, representing exchange potential as $V_x(x) = -\frac{8N^{1/2}\exp(-x^2/2b^2)}{\pi^{3/4}(bg_s)^{1/2}} \approx -\frac{8N^{1/2}}{\pi^{3/4}(bg_s)^{1/2}}(1 - \frac{x^2}{2b^2})$.

For this potential we obtain $b = \frac{\pi^{1/2}}{2}\left(\frac{g_s}{2N}\right)^{1/3}$.

Also this value of $b$ is one and a half times less, than $b$ given by (10), the critical electron density varies insignificantly, and $N_c = 0.27$.

In summary, the critical electron density below which all electrons are in a spin-polarized state in a QW is found. An estimation of the value of the critical density is obtained analytically, and this estimation is in good agreement with the results of numerical calculations.

This work was supported by Russian Foundation for Basic Research.

Electronic address: *a_vas2002@mail.ru*

[1] R. Crook, J. Prance, K.J. Thomas, S.J. Chorley, I. Farrer, D.A. Ritchie, M. Pepper, C.G. Smith, Science **312,** 1359 (2006)

[2] F.E. Orlenko, S.I. Chelkak, E.V. Orlenko, G.G. Zegrya, Zh. Eskp. Teor. Fiz. **137**, 1 (2010).

[3] N.T. Bagraev, I.A. Shelykh, V.K. Ivanov, and L.E. Klyachkin, Phys. Rev. B **70**, 155315 (2004).

[4] L. Bartosch, M. Kollar, P. Kopietz, Phys. Rev. B **67**, 092403 (2003).

[5] K. Yang, Phys. Rev. Lett. 93, 066401 (2004).

[6] C. K. Wang, K. F. Berggren, Phys. Rev. B **57**, 4552 (1998).

[7] A. Ashok, R. Akis, D. Vasileska, D.K. Ferry, Mol. Simulations **31**, 797 (2005).

[8] B. Tanatar, D.M. Ceperley, Phys. Rev. B **39**, 5005 (1989).